\documentstyle[aps,prb,multicol,psfig]{revtex}
\draft

\begin{document}
\widetext

\title{Skyrmion Physics Beyond the Lowest Landau Level Approximation}
\author{V. Melik-Alaverdian$^1$, N.\ E. Bonesteel$^2$ and
G. Ortiz$^3$} \address{ $^1$Department of Physics, University of Rhode
Island, Kingston, RI 02881\\ $^2$National High Magnetic Field
Laboratory and Department of Physics, Florida State University,
Tallahassee, FL 32310\\ $^3$Theoretical Division, Los Alamos National
Laboratory, P.O. Box 1663, Los Alamos, NM 87545} \maketitle

\widetext
\begin{abstract}

The effects of Landau level mixing and finite thickness of the
two-dimensional electron gas on the relative stability of skyrmion and
single spin-flip excitations at Landau level filling factor $\nu=1$
have been investigated.  Landau level mixing is studied by fixed-phase
diffusion Monte Carlo and finite thickness is included by modifying
the effective Coulomb interaction.  Both Landau level mixing and
finite thickness lower skyrmion excitation energies and favor
skyrmions with fewer spin flips.  However, the two effects do not work
`coherently'.  When finite thickness is included the effect of Landau
level mixing is strongly suppressed.
\end{abstract}


\begin{multicols}{2}

\narrowtext

In a seminal paper Sondhi et al.\cite{sondhi} predicted that if the
ratio of the Zeeman energy to the exchange energy is sufficiently
small the quasiparticle excitations of the $\nu=1$ integer quantum
Hall state should be charged `skyrmions' with topologically nontrivial
spin textures.\cite{sondhi} The existence of these exotic excitations
is now well confirmed by experiment.\cite{barrett,aifer,schmeller}

While the initial theoretical description of skyrmions in terms of an
effective long-wavelength nonlinear sigma model is appropriate for
excitations whose magnetization varies slowly on the scale of the
magnetic length, in typical experiments the number of reversed spins
per skyrmion is $\sim 3-4$ and a more microscopic description is
required. A step towards such a description was taken by MacDonald,
Fertig, and Brey\cite{macdonald} who introduced explicit wave
functions describing skyrmions with well defined spin quantum numbers.

In this paper we present the results of calculations in which these
microscopic wave functions have been used to study the effects of
finite thickness of the two-dimensional electron gas (2DEG) and Landau
level mixing on $\nu=1$ skyrmion states.  Finite thickness is included
by modifying the Coulomb interaction and Landau level mixing is
studied using the fixed-phase diffusion Monte Carlo
method.\cite{ortiz,melik} One motivation for this work is the
observation by Schmeller et al.\cite{schmeller} of skyrmion energy
gaps measured in transport experiments which are 25\% smaller than
theoretical estimates based on lowest Landau level
calculations,\cite{fertig} a result which led these authors to
speculate that Landau level mixing plays an important role in reducing
skyrmion excitation energies. In our calculations we find that both
finite thickness and Landau level mixing do in fact lower skyrmion
excitation energies and favor skyrmions with fewer spin flips.
However, the two effects do not work efficiently together --- the
finite thickness effect strongly suppresses the Landau level mixing
effect.

In our calculations we have used the spherical geometry\cite{haldane}
in which $N$ electrons with charge $-e$ and mass $m$ are confined to
move on the surface of a sphere of radius $R$ in the presence of a
radial magnetic field $B$.  The total number of flux quanta piercing
the surface of the sphere is then $N_{\phi} = 2 \left(R/l_0\right)^2$,
where $l_0 = \sqrt{\hbar c / e B}$ is the magnetic length, and the
Hamiltonian describing the system is
\begin{equation}
H = \frac{1}{2m} \sum_{i=1}^N \left({\hat {\bf
r}}_i\times\left(\frac{\hbar}{i}{\bf\nabla}_i+\frac{e}{c}{\bf
A}_i\right)\right)^2 + V_C + H_Z.
\end{equation}
Here $V_C = \sum_{i<j} V(r_{ij})$ is the Coulomb interaction where
$r_{ij}$ is the chord distance between electrons $i$ and $j$ and $H_Z
= - g \mu_B \sum_i {S_i}_z B$ is the Zeeman interaction.  We work in
the Wu-Yang gauge in which the vector potential is ${\bf A} = {\bf
e}_\phi \hbar c N_\phi (1-\cos\theta)/(2 e R \sin\theta).$\cite{wu}
The softening of the short-range part of the Coulomb interaction due
to finite thickness of the lowest subband wave function of the 2DEG is
included through the modified interaction $V(r) = e^2/\epsilon
\sqrt{r^2 + \beta^2}$ where $\beta$ is a parameter characterizing the
thickness.\cite{zhang}

For the $\nu=1$ ground state the electron spins are fully polarized
and the lowest Landau level is completely full.  On the sphere this
occurs when $N = N_\phi+1$ and the corresponding wave function is a
Slater determinant of lowest Landau level wave functions.  If the
electron positions are given in terms of the complex coordinate $z =
\tan (\theta/2) \exp{(-i\phi)}$, where $\theta$ and $\phi$ are the
usual spherical angles, then, exploiting the Vandermonde form of the
Slater determinant, the ground state wave function can be written
\begin{eqnarray} 
\psi_{gs} &=& U^{N_{\phi}/2}
\prod_{i < j } (z_i - z_j)\otimes
(\uparrow_1\uparrow_2\cdots\uparrow_N)
\end{eqnarray}
where $U = \prod_k (1+|z_k|^2)^{-1}$.

Single spin-flip excitations of the system are constructed by removing
a spin up electron from the ground state, flipping its spin, and
reintroducing it into the lowest spin-down Landau level.  Letting $K$
denote the number of reversed spins associated with a given
quasiparticle excitation this procedure produces a $K=0$ quasihole
with charge $+e$ and a $K=1$ quasielectron with charge $-e$. The
transport gap which determines the activated temperature dependence of
the longitudinal resistivity at $\nu=1$ is set by the excitation
energy for creating a quasielectron and quasihole with infinite
separation.  On the sphere the best approximation to this state can be
realized by placing the quasihole at the bottom of the sphere ($\theta
= \pi$) and the quasielectron at the top of the sphere ($\theta = 0$).
The corresponding wave function is
\begin{eqnarray}
\psi_{sf} &=& {\cal A}~U^{N_\phi/2}\prod_{i < j, i \ne 1}
(z_i - z_j)\otimes
(\downarrow_1\uparrow_2\cdots\uparrow_N)
\end{eqnarray} 
where the operator ${\cal A}$ antisymmetrizes the wave function under
the exchange of all pairs of electrons.

The nature of the quasiparticle excitations at $\nu=1$ depends on the
relative size of the Zeeman energy and the exchange energy
characterized by the dimensionless ratio $\tilde g = g\mu_B
B/(e^2/\epsilon l_0)$.  In the limit of large $\tilde g$ the transport
gap for the $\nu=1$ integer quantum Hall state is set by the single
spin-flip excitation described above. However, as $\tilde g$ is
lowered below a critical value the single spin flip becomes unstable
to the formation of a neutral skyrmion pair, i.e.  a well separated
charge $+e$ and charge $-e$ skyrmion pair, involving more than one
flipped spin.\cite{sondhi} A microscopic wave function description of
the charge $+e$ skyrmion excitations was developed by MacDonald,
Fertig, and Brey\cite{macdonald} who showed that for a model
Hamiltonian with a hard-core delta-function repulsion the states
\begin{eqnarray}
\psi_{sk}(K) &=& {\cal A}~U^{(N_\phi+1)/2}
\prod_{i < j} (z_i -
z_j)\nonumber \\  
&\times& z_{K+1}~z_{K+2}\cdots z_N\otimes
(\downarrow_{1} \cdots \downarrow_{K} 
\uparrow_{K+1} \cdots \uparrow_N)
\label{hardcore}
\end{eqnarray}
are exact eigenstates of $H$ with charge $e$ and $K$ reversed spins.

The wave functions $\psi_{gs}$, $\psi_{sf}$ and $\psi_{sk}(K)$
describe states in which the kinetic energy is completely quenched and
there is no Landau level mixing.  Therefore the only contributions to
the energy gaps computed using these wave functions will come from the
Coulomb and Zeeman energies.  The variational result for the Coulomb
contribution to the single spin-flip energy gap is
\begin{eqnarray}
\Delta_{sf}= \langle \psi_{sf}|V_C| \psi_{sf}\rangle - \langle
\psi_{gs}|V_C| \psi_{gs}\rangle.
\end{eqnarray}
In the absence of Landau level mixing an exact particle-hole symmetry
in the lowest Landau level transforms a charge $+e$ skyrmion with $K$
reversed spins into a charge $-e$ skyrmion with $K+1$ reversed spins.
It is therefore possible to compute the energy gap for creating a
neutral skyrmion pair by first computing $\Delta_{sf}$ and then
computing the change in Coulomb energy of the charge $+e$ skyrmion as
the number of reversed spins is increased,
\begin{eqnarray}
\delta(K) = \langle \psi_{sk}(K)|V_C| \psi_{sk}(K)\rangle - \langle
\psi_{sk}(0)|V_C| \psi_{sk}(0)\rangle.
\end{eqnarray}
As pointed out by Abolfath et al.\cite{abolfath} the wave functions
(\ref{hardcore}) involve power-law tails which lead to strong 
finite-size effects.  In our variational calculations we have computed
$\delta(K)$ for $\beta = 0$ and $\beta = l_0$ by doing a careful
extrapolation to the thermodynamic limit, considering systems with up
to 100 electrons. For $\beta = 0$ our results agree with previous
calculations of $\delta(K)$.\cite{abolfath,kamilla}

\begin{figure}[h]
\centerline{\psfig{figure=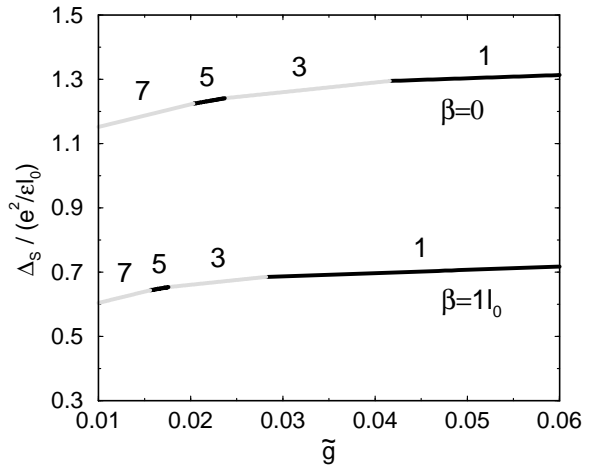,height=2.5in,angle=270}}
\vskip .15in
\caption{Energy gaps for creating neutral skyrmion pairs plotted
vs. $\tilde g$.  The number labelling each line segment is the total
number of reversed spins $S = 2K+1$. Results are given for thickness
parameter $\beta = 0$ and $\beta = l_0$ and have been extrapolated to
the thermodynamic limit.}
\label{gapvar}
\end{figure}

When the Zeeman contribution is included the total energy gap for
creating a neutral skyrmion pair for which the total number of
reversed spins $S=2K+1$ is
\begin{eqnarray}
\Delta_S(\tilde g) = \Delta_{sf} + 2 \delta(K) + S \tilde g \frac{e^2}{\epsilon
l_0}.
\end{eqnarray}
These gaps are plotted vs. $\tilde g$ in Fig.~\ref{gapvar} for
$S=1,3,5$ and 7 and for $\beta = 0$ and $\beta = l_0$. In contrast to
Hartree-Fock calculations\cite{fertig} which describe continuously
varying spin textures, our results show the quantum nature of the
skyrmion.  As $\tilde g$ decreases the energy gap undergoes level
crossings in which the total number of reversed spins of the neutral
skyrmion pair jumps by two.  Note that although we have extended the
$S=7$ line down to $\tilde g = 0.01$ we expect that as $\tilde g$
decreases the density of level crossings will increase until the
energy gap becomes, effectively, a smooth function of $\tilde g$.

Increasing the thickness of the 2DEG is seen to lower the skyrmion
energy gaps and favors excitations with fewer spin flips.  This effect
can be understood qualitatively by noting that as the number of
reversed spins $K$ increases the charge of the skyrmion spreads out
resulting in a lowering of its Coulomb energy.  When the thickness is
increased, and consequently the short-range part of the
electron-electron repulsion is decreased, the excitation energies of
the larger skyrmions will decrease {\it less} than those of the
smaller skyrmions.

We now turn to the effect of Landau level mixing on the skyrmion
energy gaps.  The importance of Landau level mixing in a given system
is characterized by the electron gas parameter $r_s = 1/(a_B\sqrt{\pi
n})$, where $n$ is the carrier density and $a_B = \epsilon \hbar^2/m
e^2$ is the effective Bohr radius.  It is straightforward to show that
$r_s = (\nu/2)^{1/2} (e^2/\epsilon l_0)/\hbar\omega_c$ where
$\hbar\omega_c = \hbar eB/mc$ is the cyclotron energy so that, at
fixed $\nu$, in the limit $r_s \rightarrow 0$ the cyclotron energy is
much larger than the Coulomb energy and there is no Landau level
mixing. However, in typical experiments $r_s$ is of order 1 or higher
and Landau level mixing cannot be ignored.

To study Landau level mixing we have used the fixed-phase diffusion
Monte Carlo method,\cite{ortiz} which has proven to be a useful tool
for studying the effect of Landau level mixing on quantum Hall
states.\cite{ortiz,melik} All of the wave functions considered here
have been given in the form
\begin{eqnarray}
\psi = {\cal A}~ \psi_{\rm space}({\bf r}_1,\cdots,{\bf r}_N)
\otimes(\downarrow_1\cdots\downarrow_K \uparrow_{K+1}\cdots\uparrow_N)
\label{generic}
\end{eqnarray}
where $\psi_{\rm space}$ is completely antisymmetric under the
exchange of any pair of electrons $i$ and $j$ where either $1 \le i,j
\le K$ or $K+1 \le i,j \le N$.  In a fixed-phase diffusion Monte Carlo
simulation these wave functions are used as trial states by first
writing the space part of the wave function in the form
\begin{eqnarray}
\psi_{\rm space}({\bf r}_1,\cdots,{\bf r}_N) = |\psi_{\rm space}({\bf
r}_1,\cdots,{\bf r}_N)| e^{i\varphi({\bf r}_1,\cdots,{\bf r}_N)},
\end{eqnarray}
then `fixing' the trial phase $\varphi$, and finally constructing an
effective bosonic Schr\"odinger equation for the positive definite
wave function $|\psi_{\rm space}|$ and solving that equation using
standard diffusion Monte Carlo techniques.\cite{ceperley} The result
of this procedure is the lowest energy state of the form
(\ref{generic}) subject to the constraint that the phase is the same
as that of the trial function.

\begin{figure}[h]
\centerline{\psfig{figure=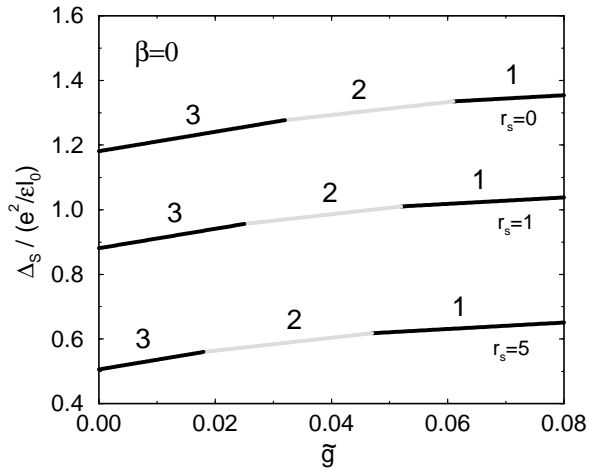,height=2.5in,angle=270}}
\vskip .15in
\caption{Energy gaps for creating neutral excitations consisting of a
charge $+e$ single spin-flip quasielectron and a charge $-e$ skyrmion
with $K = 0,1$ and 2 reversed spins for $r_s = 0,1$ and 5 as a
function of $\tilde g$ for thickness parameter $\beta = 0$. Each line
segment is labelled by the total number of reversed spins $S = K+1$.
Results are for a 30 electron system.}
\label{rsb0}
\end{figure}

We have implemented this procedure using $\psi_{gs}$, $\psi_{sf}$
and $\psi_{sk}(K)$ as trial wave functions to obtain, respectively,
the fixed-phase energies $E^{FP}_{gs}$, $E^{FP}_{sf}$ and
$E_{sk}^{FP}(K)$ for a system with 30 electrons.  For finite $r_s$ it
is no longer possible to transform a charge $+e$ skyrmion into a
charge $-e$ skyrmion by particle-hole symmetry. Therefore, in order to
calculate a physical quantity, we have computed the energy gaps for
creating a charge $+e$ skyrmion with $K$ spin flips and a single 
spin-flip quasielectron.  As before, letting $\Delta_{sf} = E^{FP}_{sf}
- E^{FP}_{gs}$ and $\delta(K) = E_{sk}^{FP}(K) - E_{sk}^{FP}(0)$, the
corresponding energy gaps are given by
\begin{eqnarray}
\Delta_S(\tilde g) = \Delta_{sf} + \delta(K) + S \tilde g \frac{e^2}{\epsilon l_0}.
\end{eqnarray}
where now $S = K+1$.

\begin{figure}[t]
\centerline{\psfig{figure=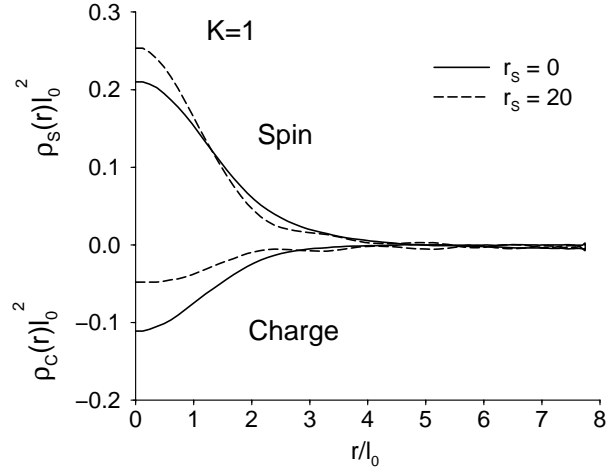,height=2.5in,angle=270}}
\vspace{.15in}
\caption{Spin and charge densities as a function of $r/l_0$ ($r$ is
the chord distance on the sphere from the center of the skyrmion) for
a charge $+e$ skyrmion with $K=1$ for $r_s= 0$ and $r_s = 20$.
Results are for a 30 electron system.}
\label{densityrs}
\end{figure}

These energy gaps are plotted as a function of $\tilde g$ in
Fig.~\ref{rsb0} for $S=1,2$ and 3 and for $r_s = 0,1$ and 5.  As $r_s$
is increased the effect of Landau level mixing is seen to be
qualitatively similar to that of finite thickness -- the skyrmion
excitation energies are lowered, with the smaller size skyrmions
having their energies lowered the most.  This effect can be understood
qualitatively by examining the effect of Landau level mixing on the
skyrmion spin and charge densities $\rho_s$ and $\rho_c$ defined by
\begin{eqnarray}
\rho_{{\scriptstyle s}\atop{\scriptstyle c}}({\bf r}) &=&
\frac{\langle\psi_{sk}(K)|(\rho_\downarrow({\bf r}) \mp
\rho_\uparrow({\bf r})) | \psi_{sk}(K)
\rangle}{\langle\psi_{sk}(K)|\psi_{sk}(K)\rangle} \pm \overline{\rho}
\end{eqnarray}
where $\rho_\sigma$ is the number density operator for spin $\sigma$
and $\overline{\rho}$ is the uniform number density far from the
skyrmion.  Fig.~\ref{densityrs} shows mixed estimates\cite{ceperley}
of $\rho_s$ and $\rho_c$ for the $K=1$ charge $+e$ skyrmion for $r_s =
0$ and $r_s = 20$.  For $r_s = 20$ the possibility of mixing in higher
Landau levels leads to a spreading out of the charge density of the
skyrmion resulting in a lowering of its Coulomb energy at the cost of
some kinetic energy.  This effect is suppressed as $K$ increases
because for larger values of $K$ the charge is already well spread out
and there is less energy to be gained by allowing the charge to spread
further.

\begin{figure}[t]
\centerline{\psfig{figure=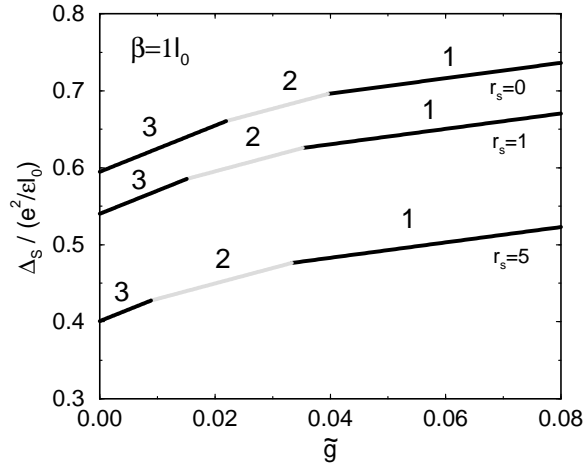,height=2.5in,angle=270}}
\vskip .15in
\caption{Energy gaps for creating neutral excitations consisting of a
charge $+e$ single spin-flip quasielectron and a charge $-e$ skyrmion
with $K = 0,1$ and 2 reversed spins for $r_s = 0,1$ and 5 as a
function of $\tilde g$ for thickness parameter $\beta = l_0$. Each
line segment is labelled by the total number of reversed spins $S =
K+1$.  Results are for a 30 electron system.}
\label{rsb1}
\end{figure}

In Fig.~\ref{rsb1} the energy gaps are shown for $S=1$, 2 and 3 and
$r_s = 0,1$ and 5 for thickness parameter $\beta = l_0$.  When finite
thickness is included the Landau level mixing effect is seen to be
much weaker than it is for zero thickness.  For example, for $\tilde g
\simeq 0.06$, when $\beta = 0$ the $r_s = 1$ energy gap is $\sim 25$\%
smaller than its $r_s = 0$ value, and when thickness is included the
$r_s=0$ energy gap drops $\sim 45\%$ for $\beta = l_0$, but the
additional reduction of the energy gap due to Landau level mixing is
only $\sim 5$\% when $r_s = 1$.

This weakening of the Landau level mixing effect due to finite
thickness can be understood as follows. The thickness correction
softens the short-range part of the Coulomb interaction, which in turn
reduces the interaction energy the skyrmion stands to gain by
delocalizing its charge.  It follows that as $\beta$ increases, the
quasielectron charge delocalizes less for a given value of $r_s$, and
so the reduction of the energy gap decreases.  Because of this
suppression of the Landau level mixing effect by finite thickness we
believe that the experimental observation by Schmeller et
al.\cite{schmeller} of skyrmion energy gaps which are significantly
smaller than theoretical estimates which ignore Landau level mixing is
more likely to be due to disorder effects, which are poorly
understood, than Landau level mixing.  Note that a similar reduction
of the Landau level mixing effect due to finite thickness has been
observed for spin-polarized Laughlin quasielectron and quasihole
excitations in the fractional quantum Hall effect.\cite{melik}

It is apparent from Figs.~2 and 4 that although finite thickness
strongly suppresses Landau level mixing, the opposite is not the case.
This is consistent with the observations of Kr\'alik et
al.\cite{kralik} who performed variational Monte Carlo calculations of
the effect of Landau level mixing and finite thickness on the $\nu=1$
single spin-flip energy gap $\Delta_{sf}$, finding that for $r_s= 1$
the inclusion of finite thickness, of roughly the same size as that
considered here, significantly reduced the energy gap.  Based on this
observation Kr\'alik et al.\ concluded that both finite thickness {\it
and} Landau level mixing contribute equally to the reduction of the
energy gap.  However, we believe that if Kr\'alik et al.\ had first
included finite thickness and {\it then} Landau level mixing they
would have observed the same strong suppression of the Landau level
mixing effect reported here.

To summarize, using skyrmion trial states recently introduced by
MacDonald, Fertig, and Brey\cite{macdonald} we have performed
variational and fixed-phase diffusion Monte Carlo calculations of the
effect of finite thickness and Landau level mixing on skyrmion
excitations at $\nu=1$. We find that both effects lower the skyrmion
excitation energies and stabilize skyrmions with fewer spin flips.
However, we also find that the two effects do not work coherently
together --- when finite thickness is included the effect of Landau
level mixing is strongly suppressed.

We thank S.P. Shukla for useful discussions.  This work was supported
in part by US DOE Grant No.\ DE-FG02-97ER45639 and NSF Grant
No. DMR-9725080.  NEB acknowledges support from the Alfred P. Sloan
foundation and GO is supported by the US DOE.

\end{multicols}

\end{document}